\documentclass{ws-procs961x669}

\newcommand{\la}{\mathcal{A}}
\newcommand{\apj}{Astrophys.\ J.}
\begin{document}

\title{Limiting effects in tori clusters}

\author{D. Pugliese and Z. Stuchl\'{\i}k}
\address{
Research Centre for  Theoretical Physics and Astrophysics\\
Institute of Physics,
  Silesian University in Opava,\\
 Bezru\v{c}ovo n\'{a}m\v{e}st\'{i} 13, CZ-74601 Opava, Czech Republic
          E-mail:d.pugliese.physics@gmail.com}
\date{\today}
\begin{abstract}
We consider  agglomerates of  misaligned tori orbiting a supermassive black hole.   The  aggregate of tilted tori is modeled as a single   orbiting configuration by  introducing  a leading function governing the distribution of toroids  (and maximum pressure points inside the disks) around  the  black hole attractor.  The orbiting clusters are composed by   geometrically thick, pressure supported,  perfect fluid tori.
This analysis  places constraints on the existence and properties of  tilted tori and more general aggregates of orbiting  disks.
 We study the  constraints on the  tori collision  emergence  and  the instability  of the  agglomerates of   tori with  general relative inclination angles, the  possible effects of the tori geometrical thickness  and on the oscillatory phenomena.
 Some notes are discussed  on the orbiting ringed structure in dependence of  the   dimensionless parameter $\xi$ representing
the (total) BH  rotational energy extracted  versus   the mass of the BH, associating  $\xi$ to the  characteristics of the accretion processes.
\end{abstract}

\bodymatter

\newcommand{\il}{~}
\newcommand{\Qa}{\mathcal{Q}}

\def\be{\begin{equation}}
\def\ee{\end{equation}}
\def\bea{\begin{eqnarray}}
\def\eea{\end{eqnarray}}
\newcommand{\aap}{A\& A}

\newcommand{\Ga}{\mathrm{G}}
\newcommand{\Mie}{\mathcal{M}}
\newcommand{\Sa}{\mathcal{\mathbf{S}}}
\newcommand{\Ca}{\mathcal{\mathbf{C}}}
\newcommand{\Ha}{\mathcal{H}}
\newcommand{\dda}{\mathcal{D}}

\newcommand{\Sie}{\mathcal{S}}

\section{Introduction}
We study  agglomerates  of tori  orbiting one central  Kerr super-massive black hole (SMBH), in two macro-configuration models: (i) as an equatorial-\textbf{R}inged \textbf{A}ccretion \textbf{D}isk (eRAD), where  the toroids equatorial  and symmetry planes     coincide with the central Kerr BH equatorial plane\cite{pugtot,ringed,open,dsystem,multy,letter,proto-jets,long,Fi-Ringed};   (ii) as  \textbf{R}inged \textbf{A}ccretion \textbf{D}isk (RAD),  where  the  agglomerate is composed by  misaligned (tilted) tori\cite{limiting,globuli,impact-jets}.

The  eRAD tori  rotation orientation is a  well defined quantity, and a couple of eRAD toroids can be  both corotating or counter-rotating  with respect to the central spinning BH, otherwise   tori can   have a  relative alternate rotation  orientation (with an inner corotating toroid and an outer counter-rotating toroid with respect to the central attractor or \emph{viceversa}).  Tori with different relative rotation orientation constitute  an interesting case from the viewpoint of the constrains posed by the eRAD or RAD model, for example    in the case of    double  tori accretion phase or also  tori collision  which   depends generally on the spin-mass ratio of the central BH.
It is clear that an orbiting tori aggregate model could  be constructed considering  a different disk model for each toroidal component.
However in this analysis we mostly  consider a Polish doughnut  (PD) model.
This is a well known geometrically thick disk model, widely used  in literature in a variety of different applications \cite{abrafra,pugtot}.  Polish doughnut shows a remarkably good fitting of  main morphological characteristics of  thick disks also  in comparison  with more refined dynamical GRHD or GRMHD models \cite{abrafra}.
RAD is a ``constraining-models",  providing  initial configurations for  dynamical (GRMHD) situations. In general  aggregates of toroidal structures orbiting one central BH attractor can  result from  different  accreting phases of the SMBHs growing, where the  infalling materials,
having diverse  angular momentum, may trace back the  BH story \cite{Violette,Dyda:2014pia,Aligetal(2013),Carmona-Loaiza:2015fqa,Lovelace:1996kx,Romanova,ringed,multy,Nixon:2013qfa,2015MNRAS.449.1251D,Bonnerot:2015ara,Aly:2015vqa}.

This  paper is structured in two parts:
In the first part, Sec.\il(\ref{Sec:erad}), we discuss the eRAD model, focusing on the model parameters and distribution of pressure and density critical points.
We introduce the concept of leading function for the agglomeration describing the tori distribution around the central attractor.

The second part, Sec.\il(\ref{Sec:RAD}), focuses  on aggregates of  misaligned tori  (RAD), discussing
explicit solutions  of its inner structure and an adapted  parametrization for the toroidal components.
\section{The \textbf{e}quatorial \textbf{ R}inged \textbf{A}ccretion \textbf{D}isks}\label{Sec:erad}
\subsection{Leading function and geodesic structure}\label{Sec:motin}
A key step  in the  modelization of the orbiting  agglomerate  is to individuate  an adapted  ``leading function",  representing the  tori location  around the central attractor.
For large  part of this  analysis,  we can identify the leading function with the definition of fluid specific angular momentum $\ell$.  Different  choices are   the agglomerate leading functions  also possibles as we see some examples in Sec.\il(\ref{Sec:magneticase})
and Sec.\il(\ref{Sec:vist-a}). In the Kerr spacetime there is
 \bea&&\label{Eq:flo-adding}
\ell\equiv\frac{L}{{E}}=-\frac{U_\phi}{U_{t}}=-\frac{g_{\phi\phi}U^\phi  +g_{\phi t} U^t }{g_{tt} U^t +g_{\phi t} U^\phi } =-\frac{g_{t\phi}+g_{\phi\phi} \Omega }{g_{tt}+g_{t\phi} \Omega },\\&&\nonumber \Omega \equiv\frac{U^\phi}{U^{t}}=-\frac{{E} g_{\phi t}+g_{tt} L}{{E} g_{\phi \phi}+g_{\phi t} L}= -\frac{g_{t\phi}+g_{tt} \ell}{g_{\phi\phi}+g_{t\phi} \ell},
\eea
where $g_{\alpha\beta}$  are metric components in the Boyer-Lindquist coordinates, $U^\alpha$ is the fluid four velocity,  $\Omega$ is the fluid relativistic angular velocity and $(E, L)$ are  constants of motions--see for example  \cite{ringed,pugtot}.
In general,  we may interpret $E$, for
timelike geodesics, as representing the total energy of the test particle
 coming from radial infinity, as measured  by  a static observer at infinity, and  $L$ as the axial component of the angular momentum  of the particle.
 (For the PD tori orbiting in a   Kerr  spacetime the set of results known as Von Zeipel theorem holds, therefore  the fluid is barotropic  and the  surfaces of constant
pressure coincide with the surfaces of  constant density.
 In these spacetimes, the family of von
Zeipel's surfaces  does not depend on the particular rotation law of the fluid,
$\Omega=\Omega(\ell)$, but on the background
spacetime only \cite{zanotti,Koz-Jar-Abr:1978:ASTRA:,M.A.Abramowicz,Chakrabarti,Chakrabarti0}.)

The leading function provides  the  distribution of the possible   maximum points of pressure and   density in the fluids surrounding the BH, which are identified as   the RAD-``rings seeds" , coincident with torus centers  $r_{center}$, and  eventually the  minimum points of pressure of  the fluids orbiting around the central attractor.
The minimum points of pressure  are associated to the cusps  $r_{\times}$ of the PD  torus Roche lobe and are regulated by  the geodesic structure of the background, composed by the marginally stable circular orbit $(r_{mso})$, the marginally bounded circular orbit $(r_{mbo})$ and the marginally  circular orbit $(r_{\gamma})$ which is also a photon orbit.  (The cusp location in the PD model,  located in the range $]r_{mbo}, r_{mso}[$,  could  be related   to  the inner edge of an  accreting  torus.)
In the RAD and eRAD models,   the location of the   maximum points of the  pressure around the attractor are interpreted as ``rings seeds", and  are   regulated by a set of  radii, associated to the geodesics structure,  located in the ``stability region" at   $r>r_{mso}$ and defined by
\bea\nonumber
&&
r_{\mathrm{(mbo)}}^{\pm}:\;\ell^{\pm}(r_{\mathrm{mbo}}^{\pm})=
 \ell^{\pm}({{r}}_{\mathrm{(mbo)}}^{\pm})\equiv \mathbf{\ell_{\mathrm{mbo}}^{\pm}},\quad\mbox{with}\quad
  r_{({\mathrm{\gamma}})}^{\pm}: \ell^{\pm}(r_{{\mathrm{\gamma}}}^{\pm})=
  \ell^{\pm}(r_{({\mathrm{\gamma}})}^{\pm})\equiv \mathbf{\ell_{{\mathrm{\gamma}}}^{\pm}},
  \\
  &&\nonumber
\mbox{where}\quad r_{\mathrm{\gamma}}^{\pm}<r_{\mathrm{mbo}}^{\pm}<r_{\mathrm{mso}}^{\pm}<
r_{\mathrm{(mbo)}}^{\pm}<
r_{({\mathrm{\gamma}})}^{\pm}
\\\label{Eq:seasn}&&
\mbox{and } \quad r_{\mathcal{M}}^{\pm}:\partial_r\partial_r\ell^{\pm}=0, \quad \mbox{with}\quad
  r_{(\Mie)}^{\pm}:\;\ell^{\pm}(r_{(\Mie)}^{\pm})=
 \ell^{\pm}(r_\Mie^\pm))\equiv \mathbf{\ell_{\Mie}^{\pm}}.
\eea
In here and in the following with $\Qa^{\pm}$ we indicate quantities $\Qa$ associated to counter-rotating $(\ell a<0)$ or  corotating $(\ell a>0)$ structures respectively, with respect to the central BH spin $a/M$.
In the following,  for any quantity $\mathbf{Q}$ and radius $r_{\bullet}$ we adopt the notation $\mathbf{Q}_{\bullet}\equiv \mathbf{Q}(r_{\bullet})$. Where more conveniently, we   use dimensionless quantities where $r\rightarrow r/M$ and $a\rightarrow a/M$.

The leading  function,  solution of $\partial_r^2\ell^{\pm}(r,\theta; a)=0$ (where $(r,\theta,\phi,t)$ are the Boyer-Lindquist  coordinates of the Kerr metric),  provides  the point of  maximum density of the  rings seeds   distribution around the BH.

\begin{figure}
\begin{center}
  \includegraphics[width=5.6cm]{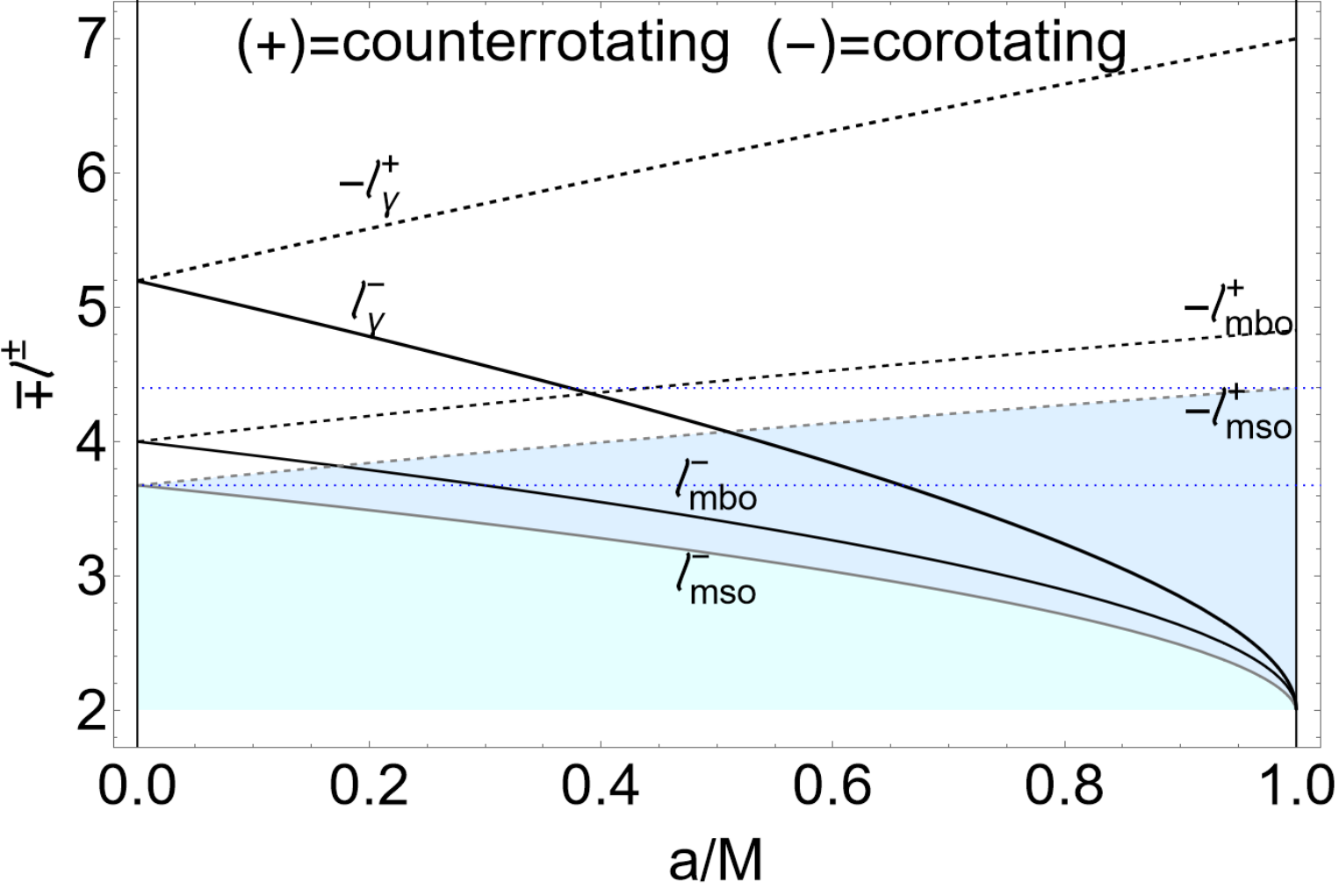}
  \includegraphics[width=5.6cm]{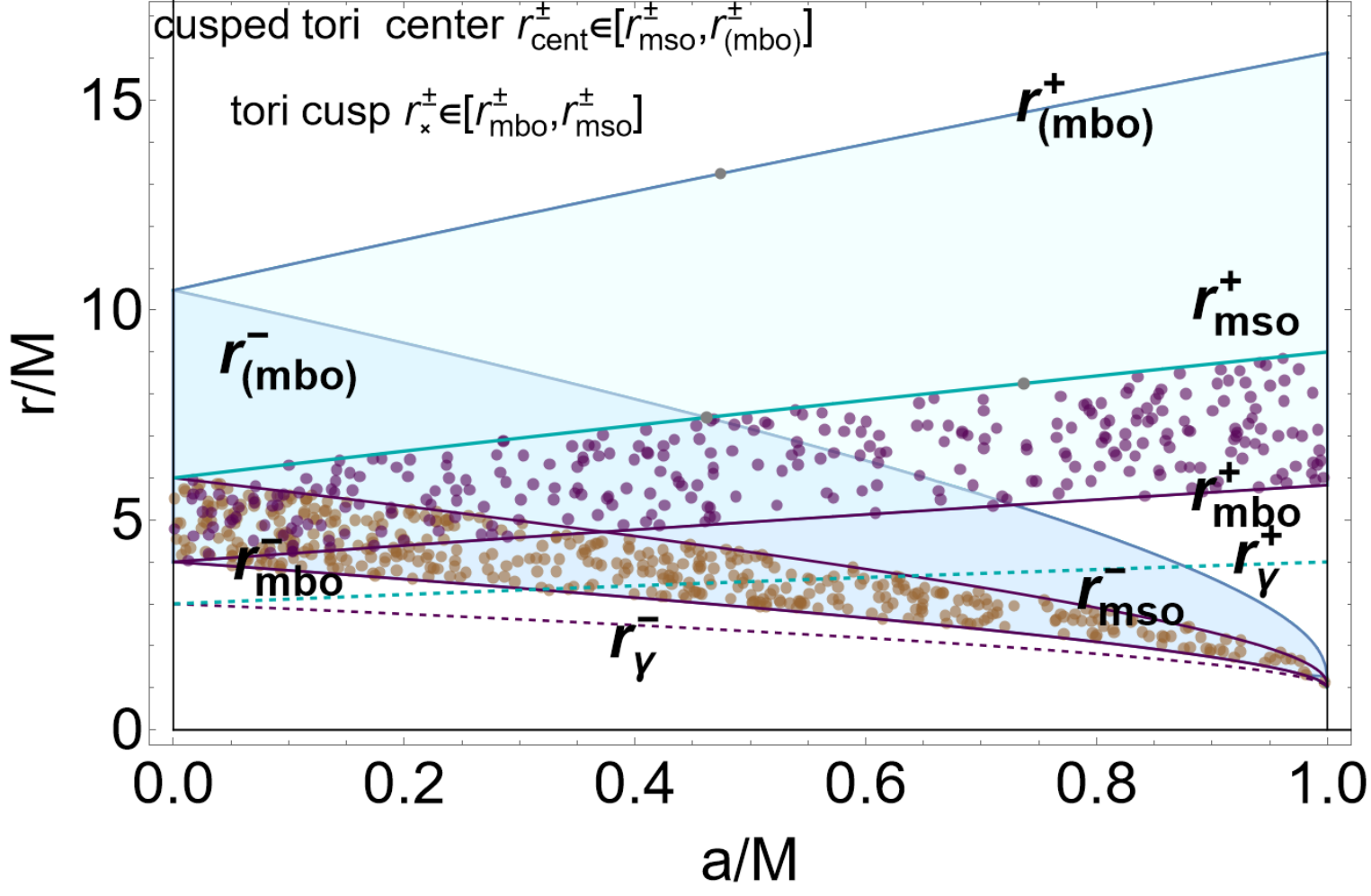}
  \end{center}
  \caption{ Left: Fluid specific angular momentum $\ell^{\pm}$ of  the {eRAD} tori, for corotating $(-)$, and counterrotating $(+)$, fluids, versus {SMBH} dimensionless spin $a/M$. Tori can form  for  $\mp\ell^{\pm}>\mp\ell_{mso}^{\pm}$ respectively. There is   $\ell_{\bullet}\equiv \ell(r_{\bullet})$ for $r_{\bullet}=\{r_{mso},r_{mbo},{r_{\gamma}}\}$, $r_{mso}$  is the marginally stable circular orbit, $r_{mbo}$ is the marginally bounded orbit and $r_{\gamma}$ the last circular (photon) orbit. Right: radii $r_{mbo}$ and $r_{mso}$ and the pair $r_{(mbo)}$ and $r_{(mso)}$ as functions of the dimensionless spin of the BH--see Eqs\il(\ref{Eq:seasn}).}\label{Fig:PlotxdisollMMb}
\end{figure}

The  corotating  and counter-rotating tori cusps orbital regions are shown  in the Figs\il\ref{Fig:PlotxdisollMMb} for different BH attractors. In the Figs\il\ref{Fig:PlotxdisollMMb}   the orbital regions for the  ring seeds locations (tori centers) are also shown. The union of these regions provides  the maximum  rage of the location of the rings seeds and  the maximum extension of the disks inner part (the region $[r_{inner}, r_{center}]$ where $r_{\times}\leq r_{inner}$ is   the torus cusp  and $r_{inner}$ is the inner edge of the quiescent (i.e. not cusped torus)). The orbital strips in  Figs\il\ref{Fig:PlotxdisollMMb}  relative to the corotating and  counter-rotating fluids, cross in different points depending on the  BH spin--mass ratio and  particularly for slower spinning attractors.
\subsection{ Ideal GRMHD and GRHD}
Polish doughnuts have been realized  in different    ideal GRMHD and GRHD setups, where  for the   ideal GRMHD (infinitely conductive plasma) case there  is
\begin{eqnarray}\nonumber
&& U_\alpha\nabla^\alpha\rho+(p+\rho)\nabla^\alpha U_\alpha=0, \label{E:1}
\\\nonumber
&& (p+\rho)U^\alpha \nabla_\alpha U^\gamma-\epsilon h^{\beta\gamma}\nabla_\beta p-\epsilon(\nabla^\alpha F_{\alpha\delta})F^{\phantom\ \delta}_\beta h^{\beta\gamma}=0,
\\\label{E:somo}\nonumber
&& U^\alpha\nabla_\alpha s=0.
\end{eqnarray}
($F_\alpha^\beta$ is the Faraday tensor and $\epsilon$ is a quantity related to the metric signature\cite{PLUS,PKroon}).
The electric field does not affect the continuity equation or the equation for the entropy.
The entropy per particles $s$ is conserved along the flow $U^\alpha$, defined for  \textit{each}  torus of the aggregate,  and $h_\alpha^\beta$ is   the metric on the 3-sheet orthogonal to  flow direction $U^\alpha$, defining  the projector tensor (there is $\nabla_\alpha g_{\beta\gamma}=0$). The  inner ringed structure is defined by the boundary conditions determining the tori (edges)---see for more details \cite{PLUS,EPL,Fi-Ringed}.

To simplify our discussion  we  consider in this analysis the GRHD scenario (see for example\cite{pugtot,abrafra}).  With the  barotropic equation of state,   $p=p(\rho)$, the set of GRHD equations for the PD model,  reduces to the only  constrain equation for the pressure (Euler equation):
 \bea\nonumber
	\frac{\nabla_\mu p}{p+\rho}=-\nabla_\mu W+\frac{\Omega \nabla_\mu \ell}{1-\Omega \ell};
\eea
where
$W=W(a; \ell; r, \theta)$ is an effective potential,  function of  $(r,\theta)$   for the Kerr metric (in the Boyer-Lindquist coordinates).
The  equation for the  pressure critical points can be further  simplified by  considering  $\ell$ as a model parameter, constant in each   torus of the aggregate.
In this  setup  the extreme  points of pressure are the extremes of the    potential $W$, and  the leading function is given by the radial derivative of the effective potential.
More precisely   the tori parameters are the couple
 $(\ell, K)$  with   $\ell=constant$ and    $V_{eff}=K=constant$ for each torus, where  $W=\ln V_{eff}$.

In this context, the eRAD leading function is
$\ell(r): \partial_r V_{eff}=0$.
%
The tori (Boyer) surfaces are  equipressure  surfaces (also  surfaces of constant  \((p,\rho, \ell, \Omega) \)) and the fluids fill every equipressure   surfaces  \cite{Boyer}.
\subsubsection{The energy function and tori energetics}
\begin{center}
    \begin{figure}
                 \includegraphics[width=5.5cm]{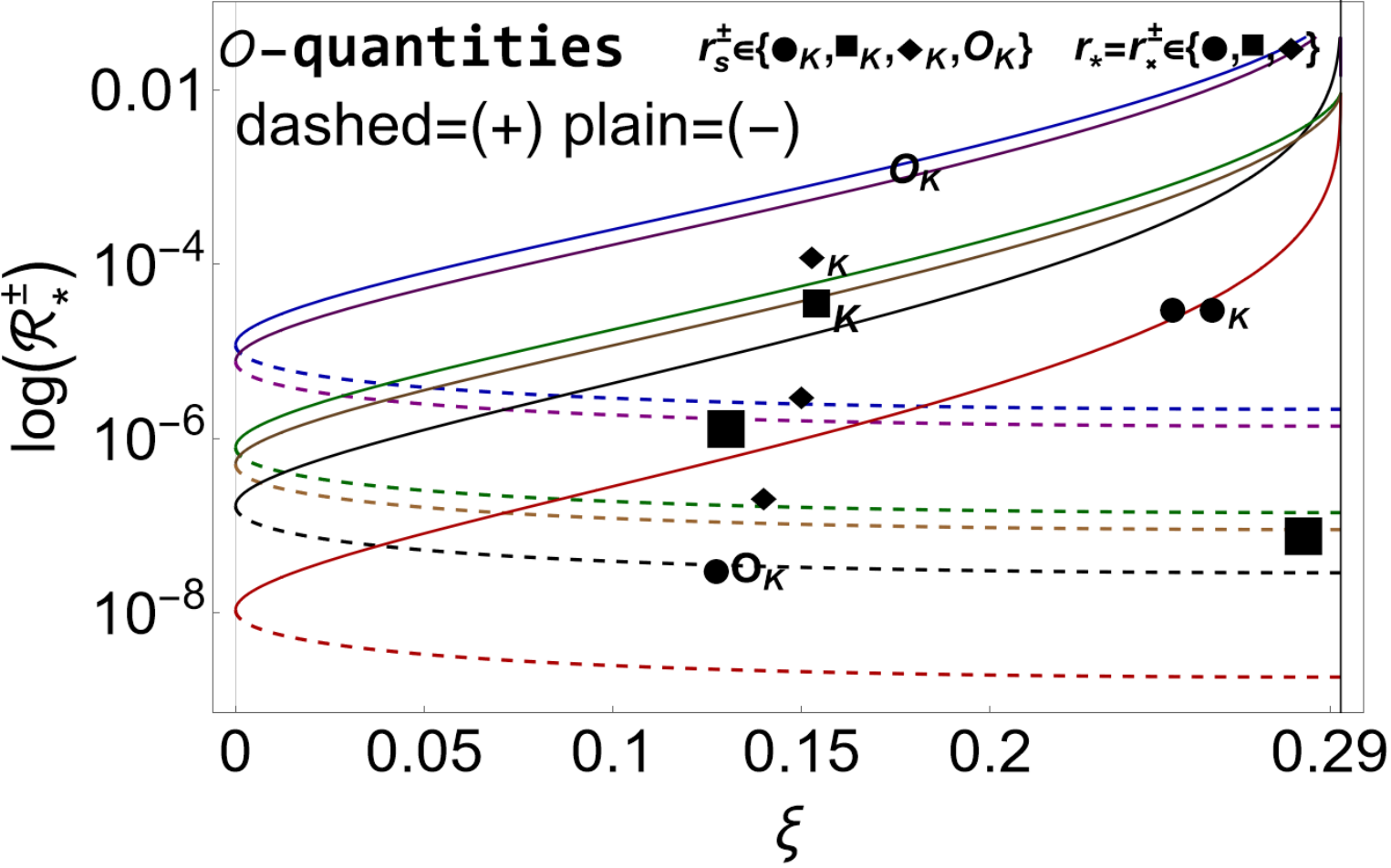}
      \includegraphics[width=5.5cm]{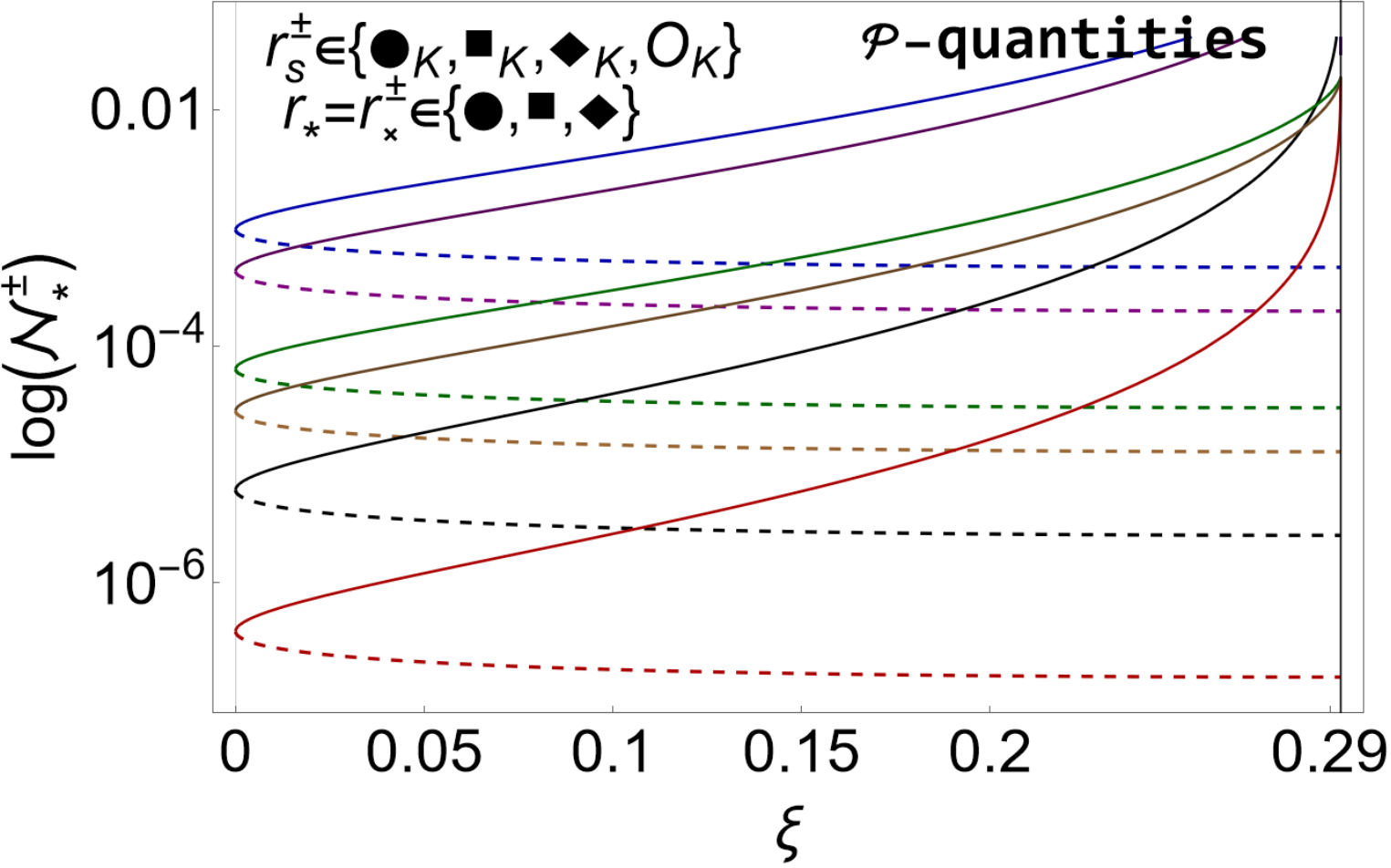}\\
        \includegraphics[width=5.5cm]{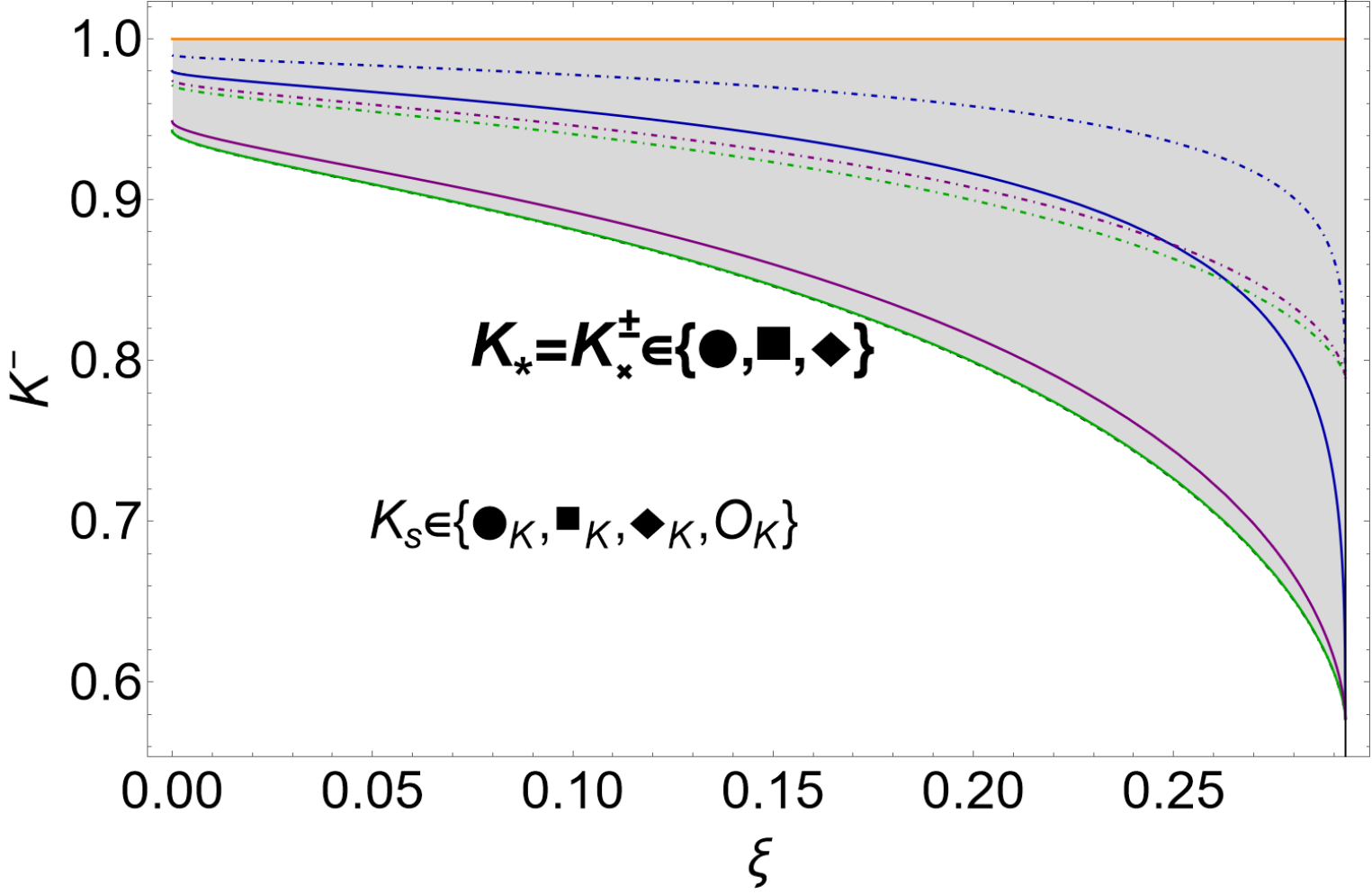}
           \includegraphics[width=5.5cm]{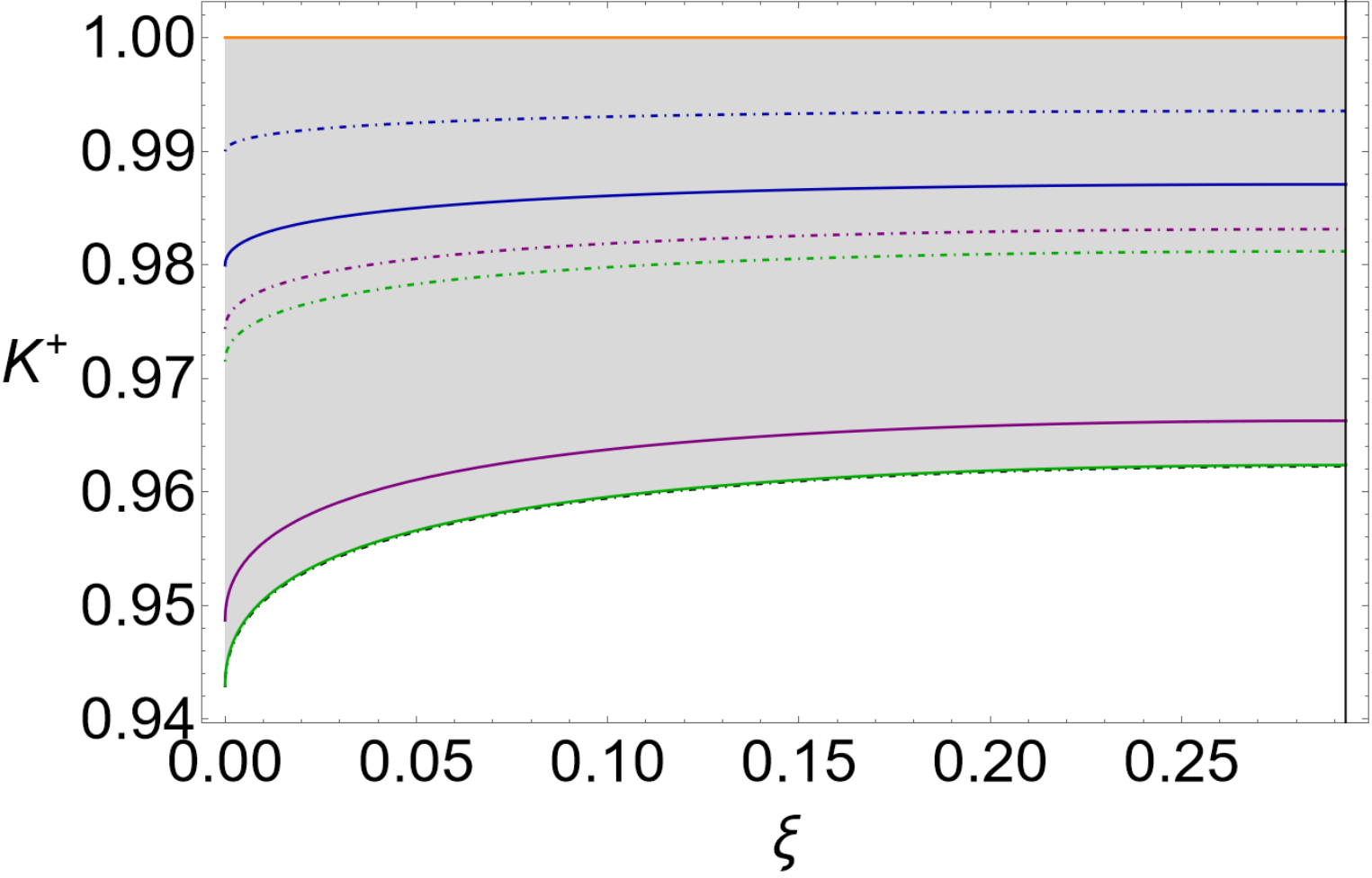}\\
                  \includegraphics[width=5.5cm]{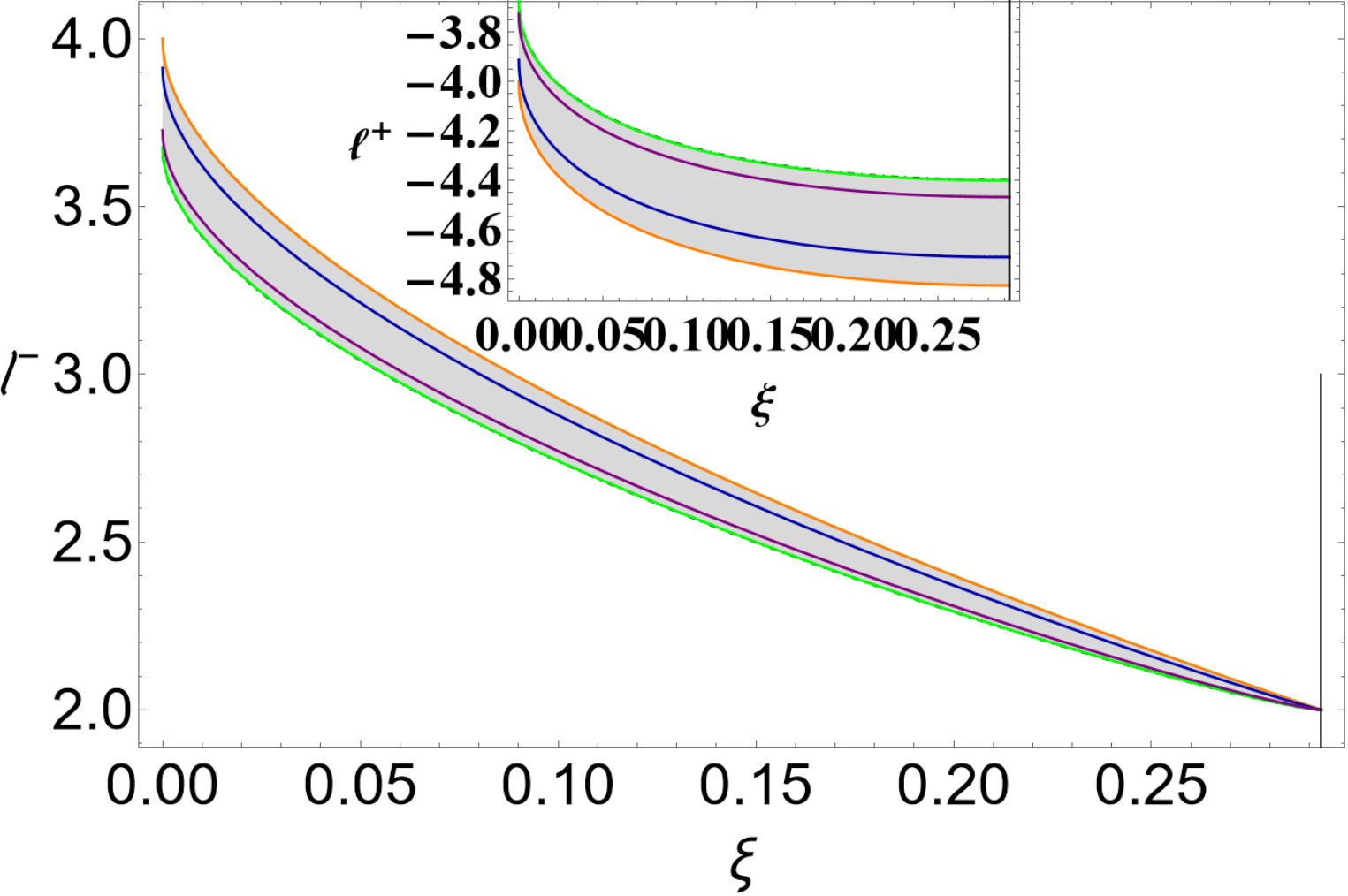}
          \includegraphics[width=5.5cm]{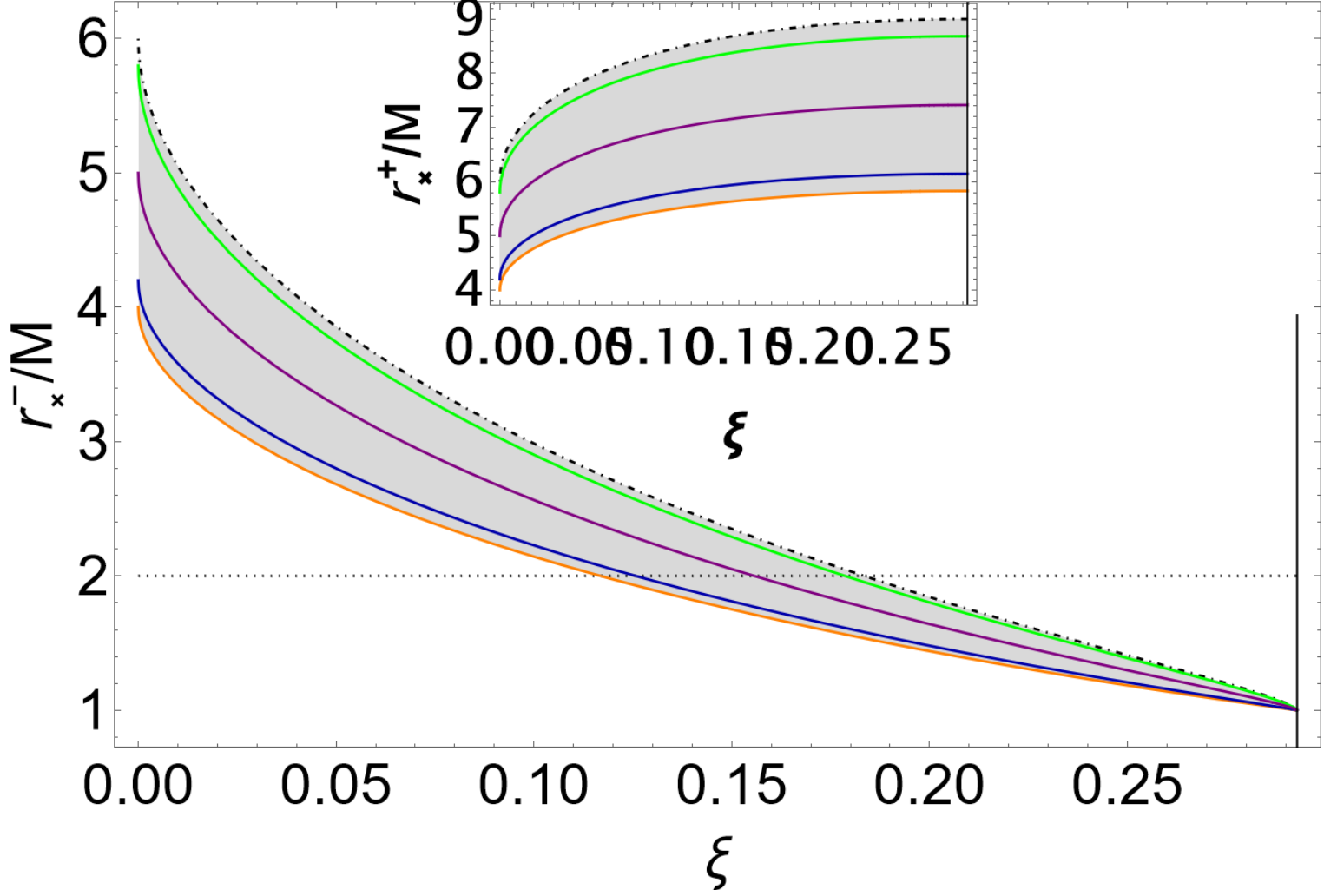}
\caption{Plots of $\mathcal{N}_{\times}^{\pm}\equiv{r_\times} (W^{\pm}(r_{s})-W^{\pm}_{\times})^\kappa(\Omega(r^{\pm}_\times))^{-1} $ for $\mathcal{P}$-quantities analysis and  $\mathcal{R}_{\times}^{\pm}\equiv(W^{\pm}(r_{s})-W^{\pm}_{\times})^\kappa$ for $\mathcal{O}$-quantities analysis defined in Table\il(\ref{Table:rock-bre}) for corotating ((-)--continuum curves) and counterrotating  ((+)--dashed curves)     tori for different values of the  cusps  $r_*=r_{\times}^{\pm}\in\{\bullet,\blacksquare,\blacklozenge\}$ and radii $r_{s}\in\{\bullet_K,\blacksquare_K,\blacklozenge_K,\mathrm{O_K}\} $,  related to thickness of the accreting  matter flow,  where  $\varpi=n+1$, with  $\gamma=1/n+1$ is the polytropic index, $\kappa$ is the polytropic constant.
 Radii $(r_\times,r_s)$ and  the associated    angular momentum  $\ell$ and $K$ parameters  are shown with $\{\bullet,\blacksquare,\blacklozenge,\bullet_K,\blacksquare_K,\blacklozenge_K,\mathrm{O_K}\}$.  $\Omega$ is  the relativistic angular  velocity. $\xi$ is the dimensionless BH rotational energy.}\label{Fig:begin-res}
\end{figure}
 \end{center}

The ``energy function"  $K(r)=V_{eff}(\ell(r))$ regulates, with the leading function $\ell(r)$, the RAD aggregate, where  $K: K(r)=$constant for each torus.
More precisely,  function
$K(r, \theta; a)\equiv \left.V_{eff}(\ell, r, \theta; a)\right|_{\ell(r,\theta;a)}$  determines   the flow (and the torus)  geometrical thickness,  the tori extension on the equatorial plane, and it is uniquely identified by $\ell(r)$ in the case of cusped tori. The relation between the geometrical maxima  (defined by $K$) and the density  maxima (fixed by $\ell$) is provided by the extreme of the leading function $\ell(r,\theta;a)$ \cite{impact-jets}. We can relate  $K(r)$ to certain features of the tori energetics, evaluating some characteristics   related to the   flow thickness, as mass accretion rate or cusp luminosity as  listed in Table\il(\ref{Table:rock-bre}). It is clear that   these quantities     depends on the details of  the  different   tori models, but this  analysis  can provide an   estimation  of  these quantities with respect to the flow co-rotation  or counter-rotation, the dimensionless BH rotational energy (or dimensionless spin) and the tori location in the aggregate \cite{Japan}.
\begin{center}
{\tablefont
\begin{table}
\tbl{ There is  $\varpi=n+1$, with  $\gamma=1/n+1$  being the polytropic index, $\kappa$ is the polytropic constant. $\Omega$ is  the relativistic angular velocity.
 $W=\ln V_{eff}$ is the value of the equipotential surface, which is taken with respect to the asymptotic value, $ W_{\times}=\ln K_{\times}$ is $W$ at  the cusp $r_\times$, while $W_s\geq W_{\times}$ and $r_s$ is related to the accreiting flow thickness.  $\mathcal{L}$ representing the total luminosity, $\dot{M}$ the total accretion rate where, for a stationary flow, $\dot{M}=\dot{M}_{\times}$,
$\eta\equiv \mathcal{L}/\dot{M}c^2$ the efficiency, $\mathcal{D}(n,\kappa), \mathcal{C}(n,\kappa), \mathcal{A}(n,\kappa), \mathcal{B}(n,\kappa)$ are functions of the polytropic index and the polytropic constant. $\mathcal{L}_{\times}/\mathcal{L}$ is the  fraction of energy produced inside the flow and not radiated through the surface but swallowed by central BH--see Figs\il\ref{Fig:begin-res}}
{\begin{tabular}{ll}
\toprule
 \mbox{\textbf{Quantities}}$\quad  \mathcal{O}(r_\times,r_s,n)\equiv q(n,\kappa)(W_s-W_{\times})^{d(n)}$ &   $\mbox{\textbf{Quantities}}\quad  \mathcal{P}\equiv \frac{\mathcal{O}(r_{\times},r_s,n) r_{\times}}{\Omega(r_{\times})}$\\\colrule
\text{\textbf{$\mathcal{R}$-quantities}:}  $\mathcal{R}_{\times}\equiv(W(r_{s})-W_\times)^\varpi
$&\text{\textbf{$\mathcal{N}$-quantities}:  } $    \mathcal{N}_{\times}\equiv\frac{{r_\times} (W(r_{s})-W_{\times})^\varpi}{\Omega(r_\times)}
$
 \\\colrule
$\mathrm{\mathbf{Enthalpy-flux:}}\mathcal{D}(n,\kappa) (W_s-W_\times)^{n+3/2},$&  $\mathbf{torus-accretion-rate:}\quad  \dot{m}= \frac{\dot{M}}{\dot{M}_{Edd}}$  \\
 $\mathrm{\mathbf{Mass-Flux:}}\quad \mathcal{C}(n,\kappa) (W_s-W_\times)^{n+1/2}$& $\textbf{Mass-accretion-rates:}\quad
\dot{M}_{\times}=\mathcal{A}(n,\kappa) r_{\times} \frac{(W_s-W_{\times})^{n+1}}{\Omega(r_{\times})}$
 \\
 $\frac{\mathcal{L}_{\times}}{\mathcal{L}}= \frac{\mathcal{B}(n,\kappa)}{\mathcal{A}(n,\kappa)} \frac{W_s-W_{\times}}{\eta c^2}$&     $\textbf{Cusp-luminosity:}\quad  \mathcal{L}_{\times}=\mathcal{B}(n,\kappa) r_{\times} \frac{(W_s-W_{\times})^{n+2}}{{\Omega(r_{\times})}}$
\\
\botrule
\end{tabular}}
\label{Table:rock-bre}
\end{table}}
\end{center}
Evaluation of these quantities defined in Table\il(\ref{Table:rock-bre}) are in Figs\il\ref{Fig:begin-res} as functions of the BH dimensionless rotational energy for different models, fixed according to selected values of the fluid specific angular momentum $\ell$ (fixing the cusp location and the center of maximum density) and the $K_s\in ]K_{\times}, 1[$, fixing the flow thickness at the cusp \cite{ellacorrelation}.
%
\subsection{The   BH rotational energy}
The BH rotational energy is related to the  BH geometrical features through its irreducible  mass $M_{irr}$.
From the  definition of irreducible mass  $
M_{irr}^2= \left(M^2+\sqrt{M^4-J^2}\right)/2$, where $M$ is the BH total (ADM) mass and  the  $J=aM$,   the dimensionless rotational energy $\xi$  is:
\bea&&
\xi\equiv\frac{M_{rot}}{M(0)}=1-\frac{\sqrt{1+\sqrt{1-\frac{J(0)^2}{M(0)^4}}}}{\sqrt{2}},\\&&\nonumber
\xi^{\mp}_{\pm}=1\pm\sqrt{\frac{r_{\mp}}{2}},
\quad
\frac{\delta M_{irr}}{M_{irr}}=\frac{\delta M-\delta J\omega_H^+}{\sqrt{M(0)^2-\frac{J(0)^2}{M(0)^2}}},
\eea
where $\xi\equiv \xi_-^+$, $M_{rot}$ is the rotational mass, $r_{\pm}$ are the outer and inner BH horizons,  $\omega_H^+$ is the BH relativistic frequency (light-like limiting circular frequency evaluated at the BH outer horizon $r_+$).
Thee BH dimensionless spin  is $a(\xi)=\la\equiv 2 \sqrt{-(\xi -2) (\xi -1)^2 \xi }$.
The rotational energy $\xi$ is governed by the constraint
$
\delta M_{irr}\geq 0$ thus $ (\delta M-\delta J \omega^+_H)\geq 0
$,  where
$\xi=1-M_{irr}/M$, with quantities evaluated at an initial state $(0)$. It is  $\xi\in[0,\xi_{\ell}]$ where $\xi_{\ell}\equiv \frac{1}{2} \left(2-\sqrt{2}\right)$, limiting  the total rotational energy extracted to a   $\approx 29\%$ of the total mass $M$ for a process leading an extreme Kerr BH to the static Schwarzschild BH.
\begin{center}
\begin{figure}
\centering
      \includegraphics[scale=.42]{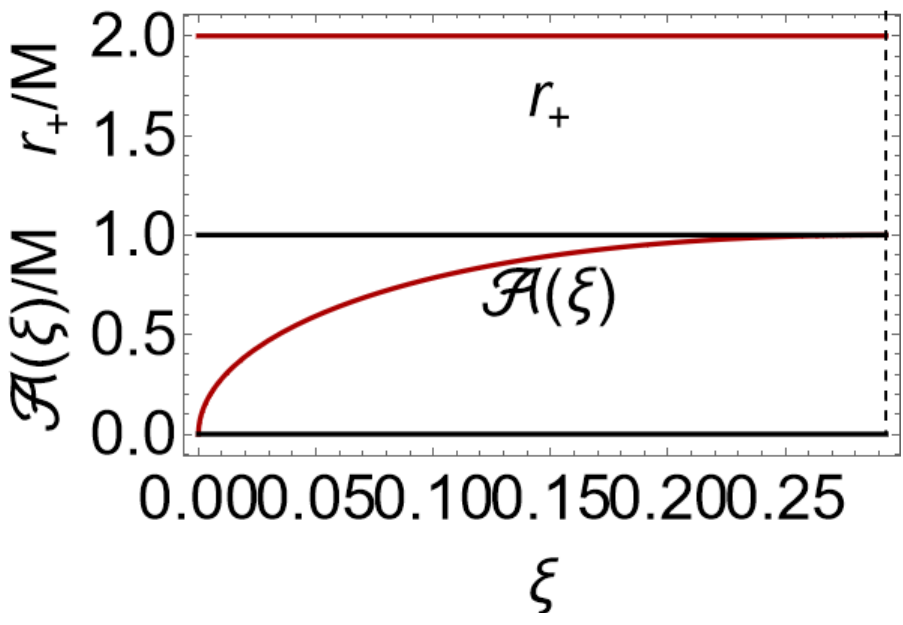}
 \includegraphics[width=4cm]{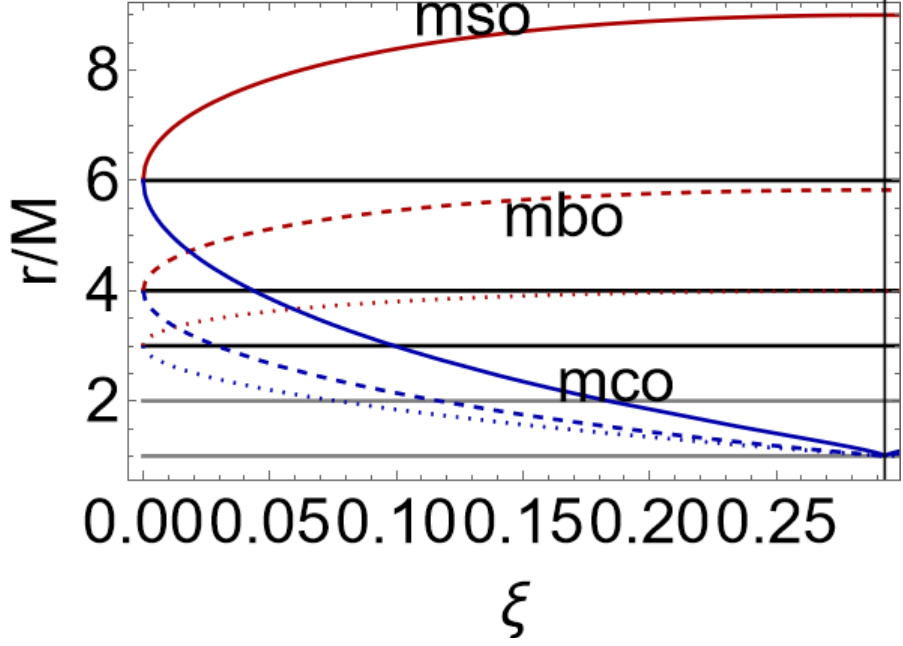}
  \\
      \includegraphics[width=4.cm]{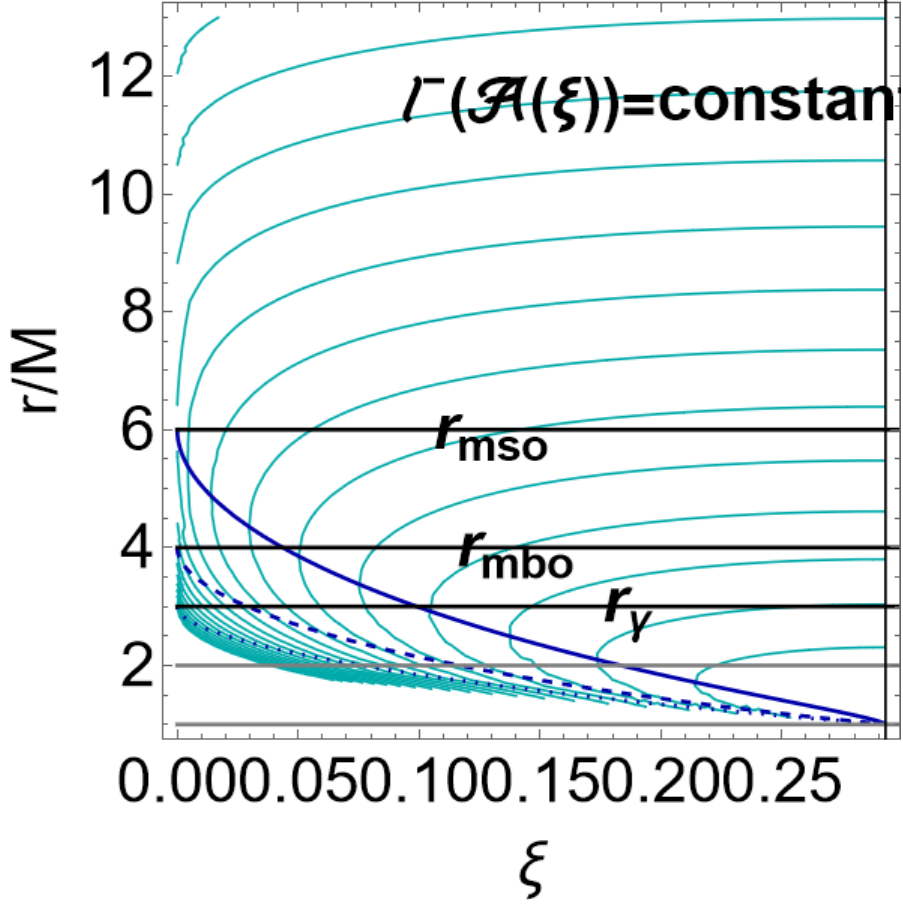}
    \includegraphics[width=4.cm]{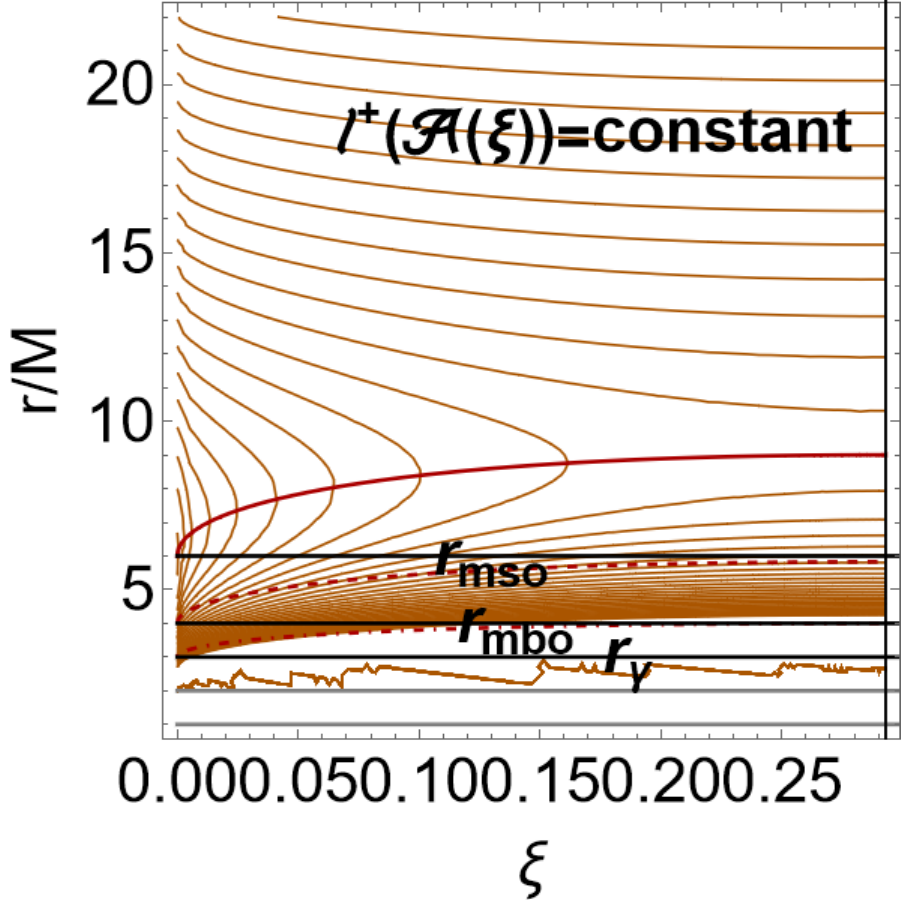}		\\
   \includegraphics[width=4.cm]{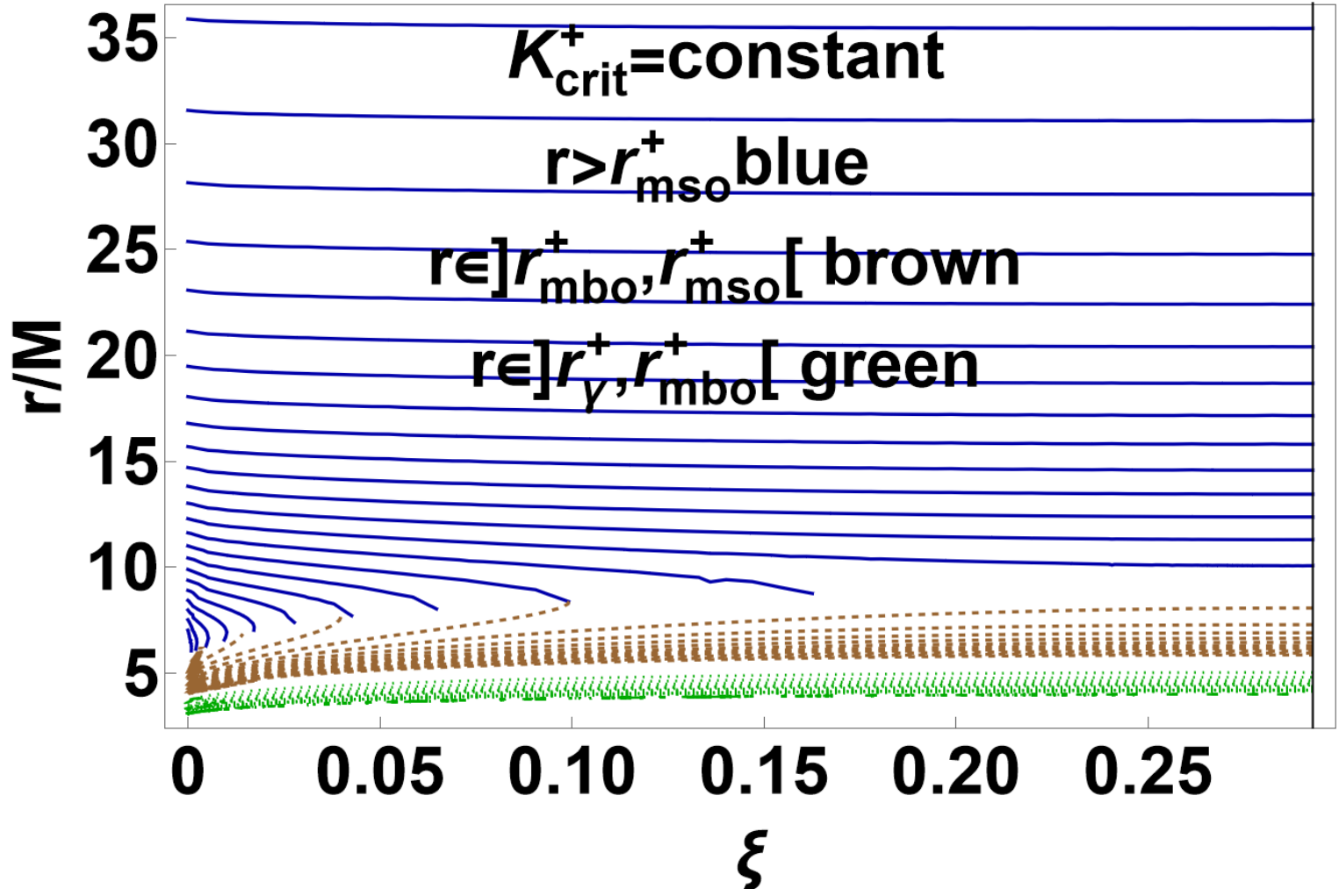}
            \includegraphics[width=4.cm]{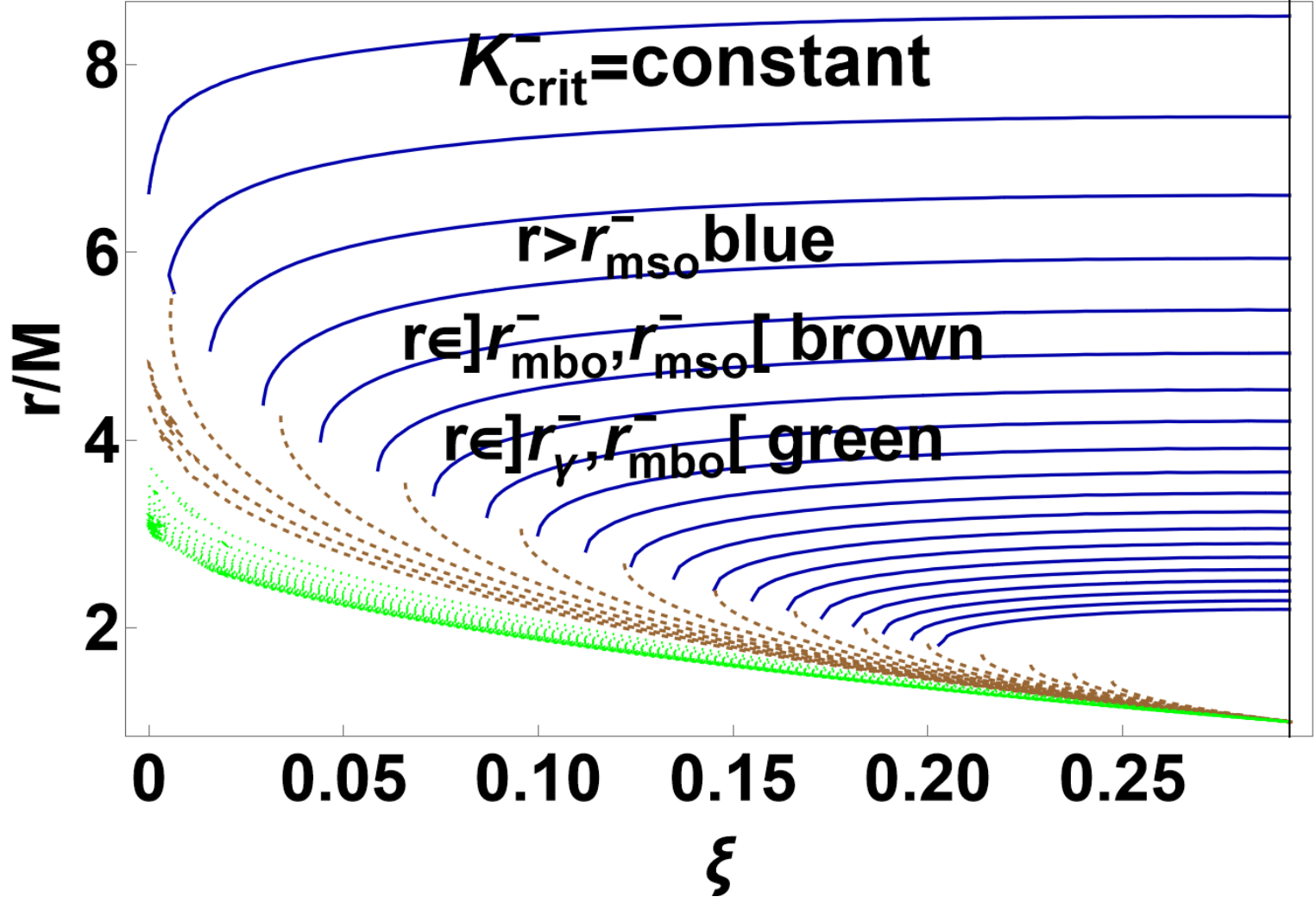}
           \caption{Upper left panel: the BH outer horizon $r_+$ and the dimenionless BH  spin  function $\la(\xi)$, as functions of the dimensionless BH rotational energy $\xi$. Extreme Kerr BH corresponds to $\xi\approx 0.29$. Upper right panel: marginally bounded orbit $mbo$ (dashed curves), marginally stable orbit $mso$ (plain curves), marginally circular orbit $mco$ (dotted curves) for corotating motion (blue curves) and counter-rotating motion (red curves) as functions of the parameter $\xi$. Center Bottom panels: the tori specific fluid angular momenta $\ell^{\mp}=$constant  for corotating (center left  panel)  and counter-rotating (center right   panel) tori, the RAD energy function $K^{\pm}(r)=$constant for the counter-rotating (bottom left panel) and  corotating fluids (bottom right panel) as functions of $\xi$, where $r_\gamma^{\pm}$ is the last circular circular orbit (photon orbit) for counter-rotating (+) and corotating orbits (-). }\label{Figs:rotW-quak}
\end{figure}
\end{center}
In Figs\il\ref{Fig:begin-res} and Figs\il\ref{Figs:rotW-quak}
  the tori aggregates  energetic characteristics  at the state $0$, prior  a possible process involving the  BH  and its environment, are related to  the  BH rotational energy. The energetic parameter $\xi$  and the angular relativistic velocity of the BH determine  the BH state prior the transition-\cite{Daly1,Daly2,ellacorrelation}.
 This analysis  turns particularly relevant for the RAD model which, because of its inner discrete structure,  can be characterized by phases of  enhanced accretion rates. Furthermore, as each torus of the aggregate  is a geometrically thick disk, each component can contribute with   super-Eddington accretion rates, with the possibility of further effects as runaway instability \cite{Abramowiczeal1983,Font:2002bi}, or in the case of RAD composed by misaligned tori, the Bardeen-Petterson effect \cite{BP75}.
  \subsection{Magnetized tori: alternative leading function}\label{Sec:magneticase}
%
%
As mentioned in Sec.\il(\ref{Sec:motin}),  the leading function is not   necessary   the fluid specific angular momentum. An example providing a different aggregate leading function is  the case of orbiting  agglomerates  composed by  magnetized tori with a ``Komissarov" toroidal magnetic field\cite{Komissarov}. In this case  the eRAD aggregate leading function  is a function  $\Sie$, defined by the magnetic field parameters and  proving the rings seeds locations,  the  minima of pressure (according to the conditions for the  cusp formation),  the maximum values  of the $\Sie$-parameter for the  formation  of a tori couple and  constrains on the eRAD inner structure in dependence on the tori relative rotation orientation.
More precisely,
let us consider the toroidal magnetic field:
$B^{\phi }=\sqrt{{2 p_B}/{{A}}}$,  where $ {A}\equiv\ell ^2 g_{tt}+2 \ell  g_{t\phi}+g_{\phi \phi }$,
and
$
p_B=\mathcal{M} \left(g_{t \phi }g_{t \phi }-g_{{tt}}g_{\phi \phi }\right){}^{q-1}\hat{\omega}^q$
is the magnetic pressure,
$\hat{\omega}$ is the fluid enthalpy, $q$  and $\Mie$ are constant;
%
The Euler equation can be written as
\bea&&
\frac{\partial_*p}{\rho+p}=G_*^{(f)}+G_*^{(em)},\quad
%
G_{*}^{\natural}=-\frac{\partial}{\partial*}W^{\natural}_{*};
\quad
*=\{r,\theta\},\quad \natural=\{(em),(f)\},
\\\nonumber
&&
 \partial_{\mu}({W^{(f)}+W^{(em)}})=\partial_{\mu}\left[\ln V_{eff}+ \mathcal{G}\right],\; \mathcal{G}(r,\theta)=\Sie \left(\mathcal{A} V_{eff}^2\right)^{q-1},\; \Sie\equiv\frac{q \mathcal{M} \hat{\omega}^{q-1}}{q-1}.
\eea
%
The  RAD leading  $\Sie$--function is
\bea\nonumber
\mathcal{\Sie}_{crit}\equiv-\frac{\Delta^{-\Qa}}{\Qa}f(a,\ell;r)\quad (Q\equiv q-1),
\eea
where $f(a,\ell;r)$ is a function of the BH spin and fluid angular momentum\cite{EPL,Fi-Ringed}
(there is $\Delta\equiv r^2-2 M r+a^2$).
\section{Misaligned tori}\label{Sec:RAD}
 In this section we discuss the  tori aggregates containing  misaligned  tori. To simply our discussion we consider misaligned tori orbiting a central static BH.
In this case we can  use different leading functions for the description of the aggregate introduced in Sec.\il(\ref{Sec:vist-a}). Tori geometrical thickness is discussed in Sec.\il(\ref{Sec:thic}). Frequency models in tori aggregates are considered  in Sec.\il(\ref{Sec: QPOs}) and
tori geometrical characteristics are deepened in Sec.\il(\ref{Sec:tecnoarr})
 for quiescent (not cusped) tori in Sec.\il(\ref{Sec:quiscne-para}) and  for cusped tori analyzed in Sec.\il(\ref{Sec:sosp-cambto}).
\subsection{Leading functions}\label{Sec:vist-a}
In this section we introduce three  definitions of leading functions for the  tori aggregates--see \cite{globuli}.

{\small \textbf{The critical $\bar{r}(r_i)$ and $r_{\times}^{\varepsilon}$ functions:}}
\bea&&\label{Eq:gocc-mas}  
\bar{r}(r_i)=\frac{2r_i \left[\sqrt{ 2 r_i-r_\gamma}+r_i-1\right]}{(r_i-r_+)^2},
\quad r_{\times}^{\varepsilon}\equiv\frac{r_{center} \left(\sqrt{2r_{center}-r_\gamma}+1\right)^2}{(r_{center}-r_+)^2},
 \eea
(solutions of $\ell(r)=\ell(r_p)$ for two orbits $(r,r_p)$).  In this case the  leading function is a relation between the extremes  of  pressure inside the tori $\bar{r}(r_i)$, or the cusp $ r_{\times}^{\varepsilon}$ as function of the center of maximum pressure. The distance $r_{center}-r_\times$  increases with the  tori distance  in the aggregates from the central  BH attractor.

\textbf{{Leading function $\ell_{crit}^o(K)$:}}

There is
 \bea\label{Eq:crititLdeval}
 &&
 \ell_{crit}^o(K)\equiv\sqrt{\frac{27 K^4-K \left(9 K^2-8\right)^{3/2}-36 K^2+8}{2K^2 \left(K^2-1\right)}},
  \\&&\ell_{crit}^i(K)\equiv\sqrt{\frac{27 K^4+K \left(9 K^2-8\right)^{3/2}-36 K^2+8}{2K^2 \left(K^2-1\right)}},
 \\\nonumber
 &&
 \mbox{where}\quad
\ell_{crit}^o(K)> \ell_{crit}^i(K)>\ell_{mso},\quad \ell_{crit}^i(K)\in[\ell_{mso},\ell_{\gamma}[.
 \eea
In this case, notably the leading function, relating the two tori parameters $\ell$ and $K$, bas been split in  function  $\ell_{crit}^o(K)$, the leading function  providing  parameters $(\ell, K)$ at the tori  centers (rings seeds), and  function $\ell_{crit}^i(K)$ for the tori cusps.

Similarly  we introduce the following alternative functions.

 \textbf{Tori critical radius  $r_{crit}^o(K)$:}
 There is
 \bea&&\nonumber
r_{crit}^o(K)\equiv -\frac{8}{K \left(\sqrt{9 K^2-8}+3 K\right)-4},\; r_{crit}^i(K)\equiv\frac{8}{K \left(\sqrt{9 K^2-8}-3 K\right)+4},\\
&& r_{crit}^i(K_{\times})=r_{\times},\; r_{crit}^o(K_{center})=r_{center}
 \eea
 similarly to  $\ell_{crit}^o(K)$,  the leading function is $r_{crit}^o(K)$, relating  the center of maximum pressure (ring seed) and the $K$ parameter at the  torus center.

%
 %
\subsection{Geometrical thickness}\label{Sec:thic}
Disk   geometrical thickness is an
 important characteristic  for the RAD  tori. The eRAD  is a geometrical thin disk composed by geometrical thick tori with an inner articulated ringed structure, combining some features of geometrical thick disks, inherited by its components and features typical of the geometrical thin disks in its global structure.  Geometrical thickness  is a relevant parameter  in the  comparison with other disks model,  in the assessment of  the torus  vertical structure and  the influence of  a possible poloidal magnetic field, for the accretion mechanism and  the study of  tori oscillations.
The definition of geometrical thickness adopted here   coincides with the  thickness $\Sa$ of the outer Roche lobe section of the PD torus.  For large part of the $(\ell, K)$ range,  cusped   tori   can be considered geometrically thin i.e. $\Sa<1$.
There are  classes of toroidal components with equal thickness. For example,   in the ``reference" case  $\Sa=1$, distinguishing geometrically thin and geometrically thick disks, where there  are couples of toroids with  equal energy parameter $K$, regulating also the flow thickness\cite{limiting,globuli}.

A further parameter  for the evaluation  of  tori geometrical thickness  is  the  dimensionless $\beta_{crit}$
\bea\label{Eq:betacrittico}\nonumber
\beta_{crit}=\frac{(r_{center}-2)^2 (r_{center}-r_{\times}) \sqrt{r_{center} r_{\times}-2 (r_{center}+2 r_{\times})}}{r_{center} \sqrt{r_{center}-3} r_{\times} \sqrt{r_{\times}-2}},
\eea
 emerging from the analysis of   cusped tori oscillation \cite{2016MNRAS.457L..19T,Straub&Sramkova(2009),Abramowiczetal.(2006)}. Similarly to the leading function
$\bar{r}(r_i)$ of Eq.\il(\ref{Eq:gocc-mas}), $\beta_{crit}$ depends on the distance between  the maximum and minimum point of pressure  in the  tori, which increases with the distance from the central attractor.
For small $\beta_{crit}$ ($\beta_{crit}\geq0$), tori may be considered geometrically thin for radial and vertical oscillation, and can be described by the
 radial and vertical epicyclical frequencies  from the hypothesis of thin (slender) tori, coincident  therefore with the respective circular orbit frequencies \cite{Straub&Sramkova(2009),G,CGU}.   The conditions for geometrical thin components according to definition $\Sa<1$ and conditions for geometrical thin tori according to $\beta_{crit}$ coincide only for  special conditions on  $\ell$ and $K$ parameters (therefore depending on the  tori location in the agglomerate and their  dimension), having tori with  combined characteristics typical of geometrical thin and thick disks\cite{limiting}.
\subsubsection{Frequency models in tori aggregates}\label{Sec: QPOs}
In the conditions where $\beta_{crit}\geq0$,  we can consider the  circular orbit approximation for  the  oscillation frequencies. In \cite{limiting}  different frequency models  are applied to the
RAD structure, interpreted  as a frame for the  high--frequency  (HF) Quasi-Periodic Oscillations (QPOs),   assuming  the geodesic  (nearly circular geodesic motion) frequencies
\bea\label{Eq:frenurnurthe}
\nu_r(r)=\nu_K(r)\sqrt{1-\frac{r_{mso}}{r}},\quad
\nu_\theta(r)=\nu_K(r)\equiv\frac{1}{r^{3/2}},
\eea
determined by the tori constraints.
The  frequencies (\ref{Eq:frenurnurthe}) are combined  for the fitting of
resonance ratios, identifying   the upper $\nu_U$ and lower $\nu_L$ frequencies from different  oscillation models and  assuming $(\nu_r(r),\nu_K(r))$  evaluated at different points $r$ of the tori surfaces.
Therefore in \cite{limiting} we used the  frequency models (\textbf{TD},\textbf{RP},\textbf{RE},\textbf{WD})  listed  Table\il(\ref{Table:models-bre}),  evaluated in different   tori models.
The   twin peak quasi-periodic oscillations with resonant frequency ratios  $\nu_U/\nu_L=\{3:2,4:3,5:4,2:1,3:1\}$ have been analyzed \cite{limiting}.
\begin{center}
{\tablefont
\begin{table}
\tbl{Frequency models and related tori models for the analysis of Sec.\il(\ref{Sec: QPOs}). Frequencies $(\nu_\theta,\nu_r)$ are  in Eq.\il(\ref{Eq:frenurnurthe}), the relativistic frequency $\nu_K=\Omega$, coincident with $\nu_\theta$ in the static spacetime is in Eq.\il(\ref{Eq:flo-adding}). The fluid specific angular momentum is $\ell$ and $K$ is the energy parameter, $r_{center}$ is the center of maximum density  and pressure in the torus, $r_{out}$ is the torus outer edge. The   relativistic-precession model  coincides in static BH   also  with the   total precession models \textbf{TP}. Radius $r_{\max}^{\times}$ is the cusped torus geometrical maximum of Eq.\il(\ref{Eqs:rssrcitt}),
where $
r_{\max}^o(r)\equiv r_{\max}^{\times}$ as functions of $\ell$.  Radii $\bar{r}(r_i)$ and $r_\times^\epsilon$ are the aggregate leading functions of Eqs\il(\ref{Eq:gocc-mas}); function $\bar{r}(r_i)$ relates the critical points of pressure in each toroidal component of the aggregate,  and  function $r_\times^\epsilon$ provides the torus cusp as function of the torus center--see Eq.\il(\ref{Eq:gocc-mas}).}
{\begin{tabular}{ll}
 \hline
\textbf{Frequency models} &   \textbf{Tori models}\\
 \toprule
\textbf{(RP)}: $\nu_U = \nu_K$,   $\nu_L = \nu_{per} \equiv \nu_K - \nu_r$; &[\textbf{(a)-model}]:   function of $r/M$
\\
(relativistic-precession model)    &[\textbf{(b)-model}]:
  $r=r_{center}(\ell)$
 function of  $\ell\in[\ell_{mso},\ell_{mbo}]$
\\
 \textbf{(RE)}:  $\nu_U =\nu_{\theta}$, $\nu_L = \nu_r;$ &[\textbf{(c)-model}]: for $r=r_{out}(\ell)$
  functions of  $\ell$
 \\
 (resonance epicyclic  models)& [\textbf{(d)-model}]:   in $r=r^{\times}_{\max}(\ell)$
  function of  $\ell$
\\
(\textbf{TD}): $\nu_L=\nu_K$,  $\nu_U = (\nu_K + \nu_r)$;&  [\textbf{(e)-model}]:
for $r=r_{\times}^{\varepsilon}$
  function of $r/M$
  torus cusp or center
\\
 (tidal distortion model)&
  [\textbf{(f)-model}]:   function of
 $\bar{r}(r_i)$
\\
(\textbf{WD}):  $\nu_L= 2 (\nu_K-\nu_r)$,  $\nu_U = (2 \nu_K- \nu_r)$; & 
[\textbf{(g)-model}]:
for $r=r_{center}(K)$  function of  $K$\\
 (warped disk model)&
\\
\botrule
\end{tabular}}
\label{Table:models-bre}
\end{table}}
\end{center}
Different components of the aggregates  fit different frequency models, according to tori  location in the aggregate with respect to the central attractor  distinguishing therefore  the toroidal components and the different torus active parts \cite{limiting}.
In the models of  Table\il(\ref{Table:models-bre}), the torus inner edge  has been considered  the active part of the emission process, the frequencies being evaluated at $r_{inner}>r_\times$ (for quiescent tori)  or $r_{inner}=r_\times$ for cusped tori,  and as  $(\beta_{crit}\gtrapprox0, \Sa\ll 1)$, the maximum of pressure point, the outer edge and the geometrical maximum point have been also considered\cite{2017AcA....67..181S,
2007A&A...463..807S,
2005A&A...437..775T,
2017A&A...607A..69K,
2016ApJ...833..273T,
2015A&A...578A..90S,
2011A&A...531A..59T,
2011A&A...525A..82S,
2008CQGra..25v5016K,
2007A&A...470..401S,
2013A&A...552A..10S,
2016A&A...586A.130S}.



\subsection{Tori geometrical characteristics}\label{Sec:tecnoarr}
Frequencies models of Table\il(\ref{Table:models-bre}) have been evaluated
 on the outer $r_{out}$  and the inner  $r_{\times}$  tori edges, the  tori geometrical maximum  $r_{\max}$ and tori center $r_{center}$.
 The evaluation of the tori geometrical characteristics is  relevant in the determination of inner ringed structure and  tori collision.
In this section we  provide $(r_{\times},r_{out},r_{center},r_{\max})$, the torus  height $h$ and the inner Roche lobe  maximum high. These quantities are  functions of  tori parameters $(\ell,K)$ for quiescent (not cusped) tori, considered in Sec.\il(\ref{Sec:quiscne-para}) and $\ell$ \emph{or} $K$ (or alternately the critical pressure points $r_\times$ and $r_{center}$) for cusped tori analyzed in Sec.\il(\ref{Sec:sosp-cambto}).
\subsubsection{Roche lobes in quiescent and cusped  tori}\label{Sec:quiscne-para}
For  quiescent and cusped tori, we provide below the  outer  and inner torus edge and the tori elongations $\lambda$ on the  tori symmetry  plane--see \cite{globuli}
\bea\label{Eq:outer-inner-l-A1}
&&
\mbox{\textbf{Torus outer edge:}}\quad
r_{out}\equiv \frac{2 \left[1+\mathbb{K} \tau  \cos \left(\frac{1}{3} \cos ^{-1}(\alpha )\right)\right]}{3 \mathbb{K}},
 \\\nonumber
 &&\mbox{\textbf{Tori elongation:}}\quad
\lambda\equiv \frac{2 \tau  \cos \left(\frac{1}{6} \left[2 \cos ^{-1}(\alpha )+\pi \right]\right)}{\sqrt{3}},\\
\nonumber&&\mbox{\textbf{Torus inner  edge:}} \quad r_{inner}\equiv\frac{2 \left[1-\mathbb{K}\tau  \sin \left(\frac{1}{3} \sin ^{-1}(\alpha )\right)\right]}{3\mathbb{K}}.
\eea
where $\mathbb{K}:\; K\equiv\sqrt{1-\mathbb{K}}$, $\Qa\equiv \ell^2$  and
 $(\alpha,\mathbb{K},\tau)$ are functions of $(\ell, K)$--see \cite{globuli}. (Note in the case of cusped tori $r_{inner}=r_{\times}$).

For the  inner Roche lobe, the inner edge an the elongation of the lobe on the symmetry  plane are
\bea&&\nonumber
 r_{inner}^{BH}\equiv\frac{2\left[\frac{1}{\mathbb{K}}-\tau  \sin \left(\frac{1}{6} \left[2 \cos ^{-1}(\alpha )+\pi \right]\right)\right]}{3},
		\\&&
		\lambda_{inner}^{BH}\equiv \frac{2}{3} \tau  \left[\sin \left(\frac{1}{6} \left[2 \cos ^{-1}(\alpha )+\pi \right]\right)-\sin \left[\frac{1}{3} \sin ^{-1}(\alpha )\right]\right].
\eea
%

The geometrical   maximum for the outer $r_{\max}^o$ and inner Roche lobes $r_{\max}^i$ are
\bea\nonumber\label{Eq:r-quest-v}
&&
\mbox{Outer lobe}:\quad
r_{\max}^o(K,\ell)\equiv\sqrt{\frac{K^2 \Qa}{K^2-1}+4 \sqrt{\frac{2}{3}} \psi  \cos \left[\frac{1}{3} \cos ^{-1}( \psi_\pi)\right]},
 \\\nonumber
&&\mbox{Inner lobe}:\quad  r_{\max}^i(K,\ell)\equiv\sqrt{\frac{K^2 \Qa}{K^2-1}-4 \sqrt{\frac{2}{3}} \psi  \sin \left[\frac{1}{3} \sin ^{-1}(\psi_\pi)\right]},
 \eea
 and the  torus height is
 \bea\nonumber
&&
h_{\max}^o(K,\ell)\equiv
\sqrt{\frac{K^2 \Qa}{1-K^2}+\mathbf{Z}-4 \sqrt{\frac{2}{3}} \psi  \cos \left[\frac{1}{3} \cos ^{-1}(\psi_\pi )\right]},
\eea
where $(\psi,\psi_\pi,\mathbf{Z})$ are functions of $(\ell, K)$ \cite{globuli}.
\subsubsection{Cusped   tori}\label{Sec:sosp-cambto}
In this section we specialize the analysis of  Sec.\il(\ref{Sec:quiscne-para}) for  cusped misaligned tori, which are  described by one only independent parameter $K$ or $\ell$ (or equivalently $r_\times$ or $r_{center}$).

We can express the critical points of pressure in the tori  in terms of the parameter $\ell$.
The  torus center  and the point of minimum density (and  hydrostatic pressure) are
\bea&&\nonumber
r_{center}(\ell)\equiv\frac{1}{3} \left[\Qa+2 L_{ \ell} \cos \left(\frac{1}{3} \iota_a\right)\right],
\quad
 r_{\times}(\ell)\equiv\frac{1}{3} \left[\Qa-2 L_{ \ell} \cos \left(\frac{1}{3} \left[\iota_a+\pi \right]\right)\right],
 \eea
where $L_\ell,L_{\mathcal{ll}}$ are functions of $\ell$ and $\Qa\equiv \ell^2$ \cite{globuli}. The cusped  torus outer edge is located at
 {\footnotesize
 \bea
&&\nonumber
r_{out}^{\times}\equiv\frac{2 \ell^2 \hat{\mathbf{\psi}}_2}{3 \ell^2 \hat{\mathbf{\psi}}_2+ \hat{\mathbf{\psi}}_0^2(6-\hat{\mathbf{\psi}}_0)}+{2 \hat{\mathbf{\psi}}_4 \sqrt{\frac{\ell^2 \left[
\hat{\mathbf{\psi}}_0^2\left(3 \ell^2\hat{\mathbf{\psi}}_2(6 -\hat{\mathbf{\psi}}_0)+\hat{\mathbf{\psi}}_0^2(\hat{\mathbf{\psi}}_0^2-12 \hat{\mathbf{\psi}}_0+36)\right)+12 \ell^2 \hat{\mathbf{\psi}}_2^2\right]}{3\left(3 \ell^2 \hat{\mathbf{\psi}}_2+ \hat{\mathbf{\psi}}_0^2(6-\hat{\mathbf{\psi}}_0)\right)^2}}},
\eea}
(where $(\hat{\psi}_0,\hat{\psi}_2)$ are functions of $\ell$).
Similarly to  functions $(\ell^{o},\ell^i)$ of Eqs\il(\ref{Eq:crititLdeval}), the leading function can be  expressed in terms of the energy function
$K$ as
{\footnotesize
\bea&&\nonumber\label{Eq:grod-lock}
K_{center}(\ell)\equiv\sqrt{\frac{\left[\Qa+2 L_{ \ell} \cos\left(\frac{ \iota_a}{3}\right)-6\right] \left[ \Qa+2 L_{ \ell} \cos \left(\frac{ \iota_a}{3} \right)\right]^2}{3 \Qa \left[3 \ell^4+2 \left(2  \Qa-15\right) L_{ \ell} \cos \left(\frac{ \iota_a}{3} \right)-39  \Qa+2 L_{ \ell}^2 \cos \left(\frac{2 \iota_a}{3} \right)+54\right]}},
 \\\nonumber
 &&K_{\times}(\ell)\equiv\sqrt{\frac{\left[ \Qa-2 L_{ \ell} \sin \left(\frac{\iota_b}{3}\right)-6\right] \left( \Qa-2 L_{ \ell} \sin \left(\frac{\iota_b}{3}\right)\right)^2}{3 \Qa \left[3 \ell^4+2 \left(15-2  \Qa\right) L_{ \ell} \sin \left[\frac{\iota_b}{3}\right]-39  \Qa-2 L_{ \ell}^2 \cos \left(\frac{2\iota_b}{3}\right)+54\right]}},
 \\
 && \iota_a\equiv \cos ^{-1}(L_{\mathcal{ll}}),\quad \iota_b\equiv  \sin ^{-1}(L_{\mathcal{ll}})
\eea
}
%
(here $L_{\mathcal{l}}$ is a function of the momentum $\ell$).
Function
$K_{center}(\ell)$ describes the  rings seeds, while $K_{\times}(\ell)$ refers the tori cusps ($\Qa\equiv \ell^2$).

The  cusped
tori outer edge  can be expressed as function of the cusp $r_\times$ as follows:
\bea&&\nonumber
r_{out}^{\times}(r_{\times})=\frac{2}{3} \left[\sqrt{\frac{(r_{\times}-r_{mso})^2 r_{\times}^2}{(r_{\times}-r_{mbo})^2}}\right.
 \left.\cos \left[\frac{1}{3} \cos ^{-1}\mathbf{X}\right]+\frac{r_{\times}}{r_{\times}-r_{mbo}}+r_\times\right]
\eea
where $\mathbf{X}$ is a function of $r_\times$,
and the
the  geometrical maxima of cusped tori for the
outer $(o)$ and inner $(i)$ Roche lobes are
\bea\nonumber\label{Eqs:rssrcitt}
&&
r_{\max}^o(r)=\sqrt{4 \sqrt{\frac{2}{3}} \psi_{\lambda } \cos \left[\frac{1}{3} \cos ^{-1}\left(-\frac{3}{4} \sqrt{\frac{3}{2}} \psi_{\lambda } \psi _{\sigma }^2\right)\right]+\frac{r^2}{(r-r_{\gamma}) \psi_{\sigma }}}
 \\\nonumber
 &&
 r_{\max}^i(r)=\sqrt{\frac{r^2}{(r-r_{\gamma}) \psi _{\sigma }}-4 \sqrt{\frac{2}{3}} \psi _{\lambda } \cos \left[\frac{1}{3} \left(\cos ^{-1}\left[-\frac{3}{4}\sqrt{\frac{3}{2}} \psi _{\lambda } \psi _{\sigma }^2\right]+\pi \right)\right]},
 \eea
(here $r$ is the  fluid pressure critical point where  $\psi_\lambda,\psi_\sigma$ are functions of $r$).

The cusped torus height is
 \bea\nonumber\label{Eq:max-prob-A1}&&
h_{\max}^o(r_{\times})=\left(-2 \sqrt{6} \sqrt{\frac{(r_{\times}-r_{\gamma}) (r_{\times}-r_+)^2 r_{\times}^4}{(r_{\times}-r_{mbo})^3}} \sec \left[\frac{1}{3} \cos ^{-1}(\psi_\rho )\right]+\right.
\\\nonumber
&&
\left.\frac{9 (r_{\times}-r_+)^2 r_{\times}^2 \sec ^2\left[\frac{1}{3} \cos ^{-1}(\psi_\rho )\right]}{8 (r_{\times}-r_{mbo}) (r_{\times}-r_{\gamma})}+\frac{(r_{\times}-r_+) (5 r_{\times}-18) r_{\times}^2}{(r_{\times}-r_{mbo})^2}\right)^{1/2},
\eea
($\psi_\sigma$ is a function of $r_\times$.)
Finally we can express the
cusped tori  geometrical thickness $\Sa_{\times}=2 h_{\times}/(\lambda_{\times})$ in terms of the pressure critical points  where $\lambda_\times$ is the cusped torus elongation on its symmetry plane and $h_\times$ the cusped torus height.

\section{Conclusion}
We explored  models of tori clusters   orbiting around  a  central SMBH, detailing  the morphological characteristics of the toroidal components.
Configurations considered here   can be used as   initial data  for  dynamical (time-dependent, evolutive)  GRMHD analysis.
A ``leading function"  as been used to  constraint the tori distribution around the central attractor, together with  the  energy function $K(r)$ regulating  the    agglomerate stability (cusp emergence and tori collision),  the flow thickness,  mass accretion rate and cusp luminosity.

From the observational viewpoint the inner ringed structure offers  several  interesting scenarios arising from the   unstable states associated to  its inner   activity, as the presence of  multiple accretion points and  inter disk shells of multiple jets.
Eventually observational evidence   of the RAD and the associated inter disk  activity could  be found in the  obscuration
of   the X-ray emission spectrum, as a track of the  agglomerate inner  composition. An indication of the presence of multiple orbiting tori could be  seen in an increasing BH accretion  mass rate and the presence of   interrupted phases of  BH accretion, or in the  emission associated to oscillation tori  modes  as in HF  QPOs.
The   establishment of  runaway  instability and  the  tori self-gravity can be  relevant further   factors for  eRAD tori agglomerate around SMBHs\cite{Abramowiczeal1983,Font:2002bi} and the  Bardeen--Petterson effect is main  relevant in the misaligned tori case \cite{BP75}.

\bibliographystyle{ws-procs961x669}

\end{document}